\documentclass[]{spie}  

\usepackage{amsmath,amsfonts,amssymb}
\usepackage{graphicx}
\usepackage[colorlinks=true, allcolors=blue]{hyperref}

\title{The XGIS instrument on-board THESEUS: Monte Carlo simulations for response, background, and sensitivity}

\author[a,b]{Riccardo~Campana}
\author[a,b]{Fabio~Fuschino}
\author[a]{Claudio~Labanti}
\author[c]{Sandro~Mereghetti}
\author[a,b]{Enrico~Virgilli}
\author[a]{Valentina~Fioretti}
\author[a]{Mauro~Orlandini}
\author[a]{John~Buchan~Stephen}
\author[a]{Lorenzo~Amati}
\affil[a]{INAF/OAS, Via Gobetti 101, I-40129, Bologna, Italy}
\affil[b]{INFN-Sezione di Bologna, Via Berti Pichat 6/7, I-40123, Bologna, Italy}
\affil[c]{INAF/IASF-Milano, Via A. Corti 12, I-20133, Milano, Italy}

\authorinfo{Further author information: Send correspondence to riccardo.campana@inaf.it}

\begin{document} 
\maketitle

\begin{abstract}
The response of the \emph{X and Gamma Imaging Spectrometer} (XGIS) instrument onboard the \emph{Transient High Energy Sky and Early Universe Surveyor} (THESEUS) mission, selected by ESA for an assessment phase in the framework of the Cosmic Vision M5 launch opportunity, has been extensively modelled with a Monte Carlo Geant-4 based software. In this paper, the expected sources of background in the Low Earth Orbit foreseen for THESEUS are described (e.g. diffuse photon backgrounds, cosmic-ray populations, Earth albedo emission) and the simulated on-board background environment and its effects on the instrumental performance is shown.
\end{abstract}


\section{INTRODUCTION}

THESEUS\footnote{\url{http://www.isdc.unige.ch/theseus/}} (\emph{Transient High Energy Sky and Early Universe Surveyor})\cite{amati20} is an ESA M5 candidate mission, currently in Phase A, designed to explore the Early Universe (the cosmic dawn and reionization eras) through a complete census of the Gamma-Ray Burst (GRB) population in the first billion years of the Universe. Moreover, the mission will study the global star formation history up to redshifts $z \gtrsim 10$, and detect the primordial population III stars, investigate the re-ionization epoch, the interstellar medium (ISM), the intergalactic medium (IGM) and the properties of the early galaxies determining their star formation properties in the re-ionization era.
 
Furthermore, THESEUS will perform a deep monitoring of the X-ray transient Universe in order to locate and identify the electromagnetic counterparts of gravitational radiation and neutrino sources, providing real-time triggers and accurate locations of GRBs and high-energy transients for follow-up with next-generation optical-NIR (E-ELT, JWST), radio (SKA), X-rays (ATHENA), TeV (CTA) telescopes.

THESEUS was selected in 2018 to carry out an assessment phase study, competing for a launch opportunity in 2032. The ESA M5 Phase A study will be completed in spring 2021 with the final selection shortly afterwards.

THESEUS is designed to carry onboard two high energy monitoring instruments (SXI and XGIS) with large field of view (FOV), and an optical/near-IR telescope (IRT) with both imaging and spectroscopic capability.
The \emph{Soft X-ray Imager}\cite{obrien20} (SXI) consists of two “Lobster-Eye” X-ray (0.3--5 keV) telescopes with CMOS detectors in the focus covering a total FOV of 0.5 sr with 0.5$'$--2$'$  source location accuracy. 
The \emph{X-Gamma ray Imaging Spectrometer}\cite{labanti20} (XGIS) consists of two units operating in the 2~keV -- 10~MeV energy range with spectrometric capabilities and operating also as imager with a total FOV of $77^\circ\times117^\circ$, overlapping the SXI one, in the 2--150~keV range. 
The \emph{Infra-Red Telescope}\cite{gotz20} (IRT) consists of a 0.7 m diameter near infrared (0.7--1.8 $\mu$m range) telescope with a focal plane assembly with an infrared detector and a  wheel with filters and dispersive elements. IRT provides imaging and moderate spectral capabilities ($R \sim 400$) in a $15'\times15' $ FOV.

\begin{figure}[htbp]
\centering
\includegraphics[width=8cm]{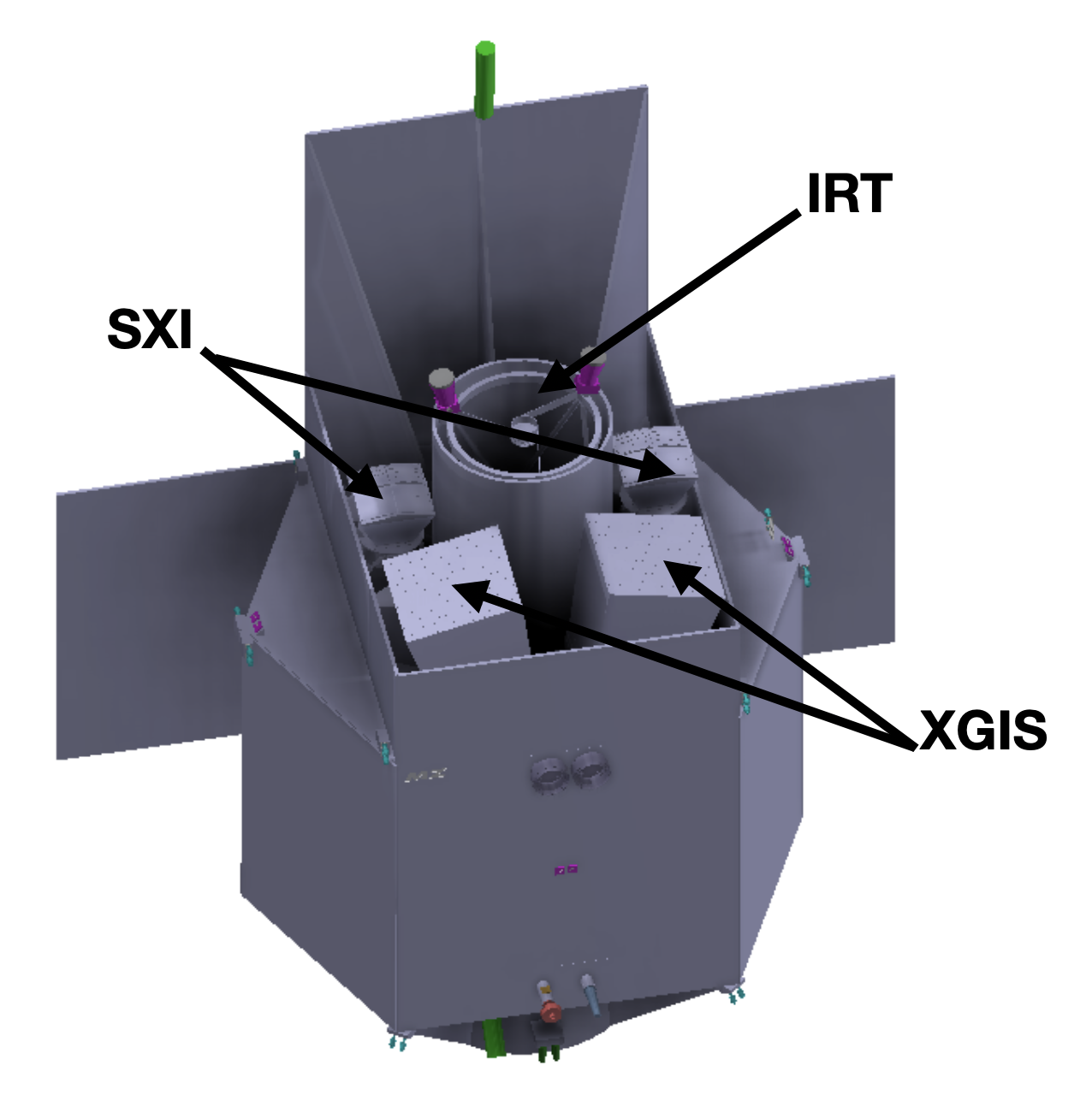}
\caption{The THESEUS mission and its instruments.}
\label{f:theseus}
\end{figure}

Figure~\ref{f:theseus} shows a sketch of the satellite configuration with its payload. The THESEUS mission is planned to be operational for at least 3 years in a circular low Earth orbit (LEO) with low inclination ($<$6$^\circ$). 

In this paper, we focus on the expected on-board background and response of the XGIS instrument (Section~\ref{s:xgis}). The expected sources of background in the Low Earth Orbit foreseen for THESEUS are described (Section~\ref{s:sources}) and the Monte Carlo mass model (Section~\ref{s:mc}) simulations of the response and background environment  are discussed (Section~\ref{s:results}).

\section{THE XGIS INSTRUMENT}\label{s:xgis}

The XGIS instrument\cite{labanti20} is both an imager up to an energy of about 150 keV (with a field of view that fully overlaps the SXI one) and a spectrometer covering a wide energy range from a few keVs to a few tens of MeV (partially overlapping the SXI energy range). 
The XGIS imaging is based on the \emph{coded mask} principle, employed successfully in different energy bands on several previous missions such as GRANAT/SIGMA, BeppoSAX/WFC, INTEGRAL/JEM-X/IBIS/SPI, RXTE/ASM, Swift/BAT. The mask shadowgram is recorded by a position sensitive detector, and can then be deconvolved into a sky image. The size of the point spread function in the sky image is determined by the ratio of the mask pixel size and the mask-to-detector distance.

The XGIS instrument sensitive plane is based on the so-called ``siswich'' detection principle\cite{marisaldi05}, which uses a large array of individual Silicon Drift Detectors (SDD) coupled to scintillating CsI(Tl) crystal bars. The SDD are used both for direct X-ray detection (X-mode, at energies below about 30 keV) and as photodetectors for the readout of scintillator optical signal (S-mode, at energies above about 20 keV). 
The two types of signals are distinguished by the use of a proper electronic acquisition chain, based on a dual readout of the individual scintillator bars, that allows  to determine also the interaction depth of the high-energy photon.

\begin{figure}[htbp]
\centering
\includegraphics[width=8cm]{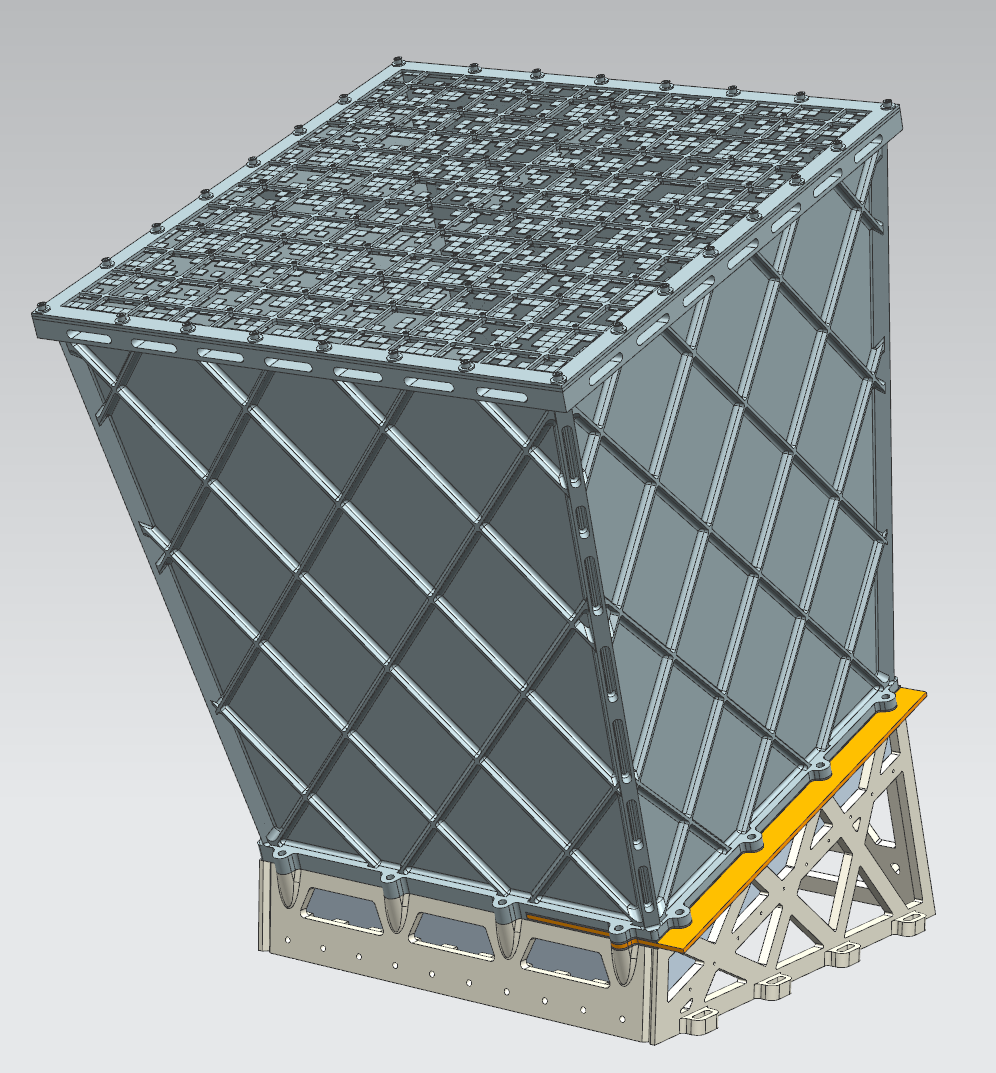}
\caption{The XGIS instrument. One unit is shown, with its mounting structure.}
\label{f:xgis}
\end{figure}

Overall, one XGIS detection unit (Figure~\ref{f:xgis}) is subdivided in 100 \emph{modules} (grouped in 10 logical and electrical \emph{super-modules}), each containing two monolithic $8\times8$ SDD arrays (i.e. each with 64 individual $5\times5$~mm$^2$ cells) coupled to 64 CsI(Tl) bars with $4.5\times 4.5\times30$~mm$^3$ dimensions.
A XGIS detection plane has therefore 6400 pixels. 

The coded mask\cite{gasent20}, placed at a distance of 63 cm, is made of tungsten. A mechanical structure (Al + W) surrounds the XGIS units ensuring the required mechanical stiffness and radiation shielding.

Two XGIS units are mounted on the optical bench of the THESEUS platform, and offset by angles of $\pm$20$^\circ$. The total field of view is therefore $77^\circ\times117^\circ$.

\section{BACKGROUND SOURCES}\label{s:sources}
\begin{figure}[htbp]
\centering
\includegraphics[width=12cm]{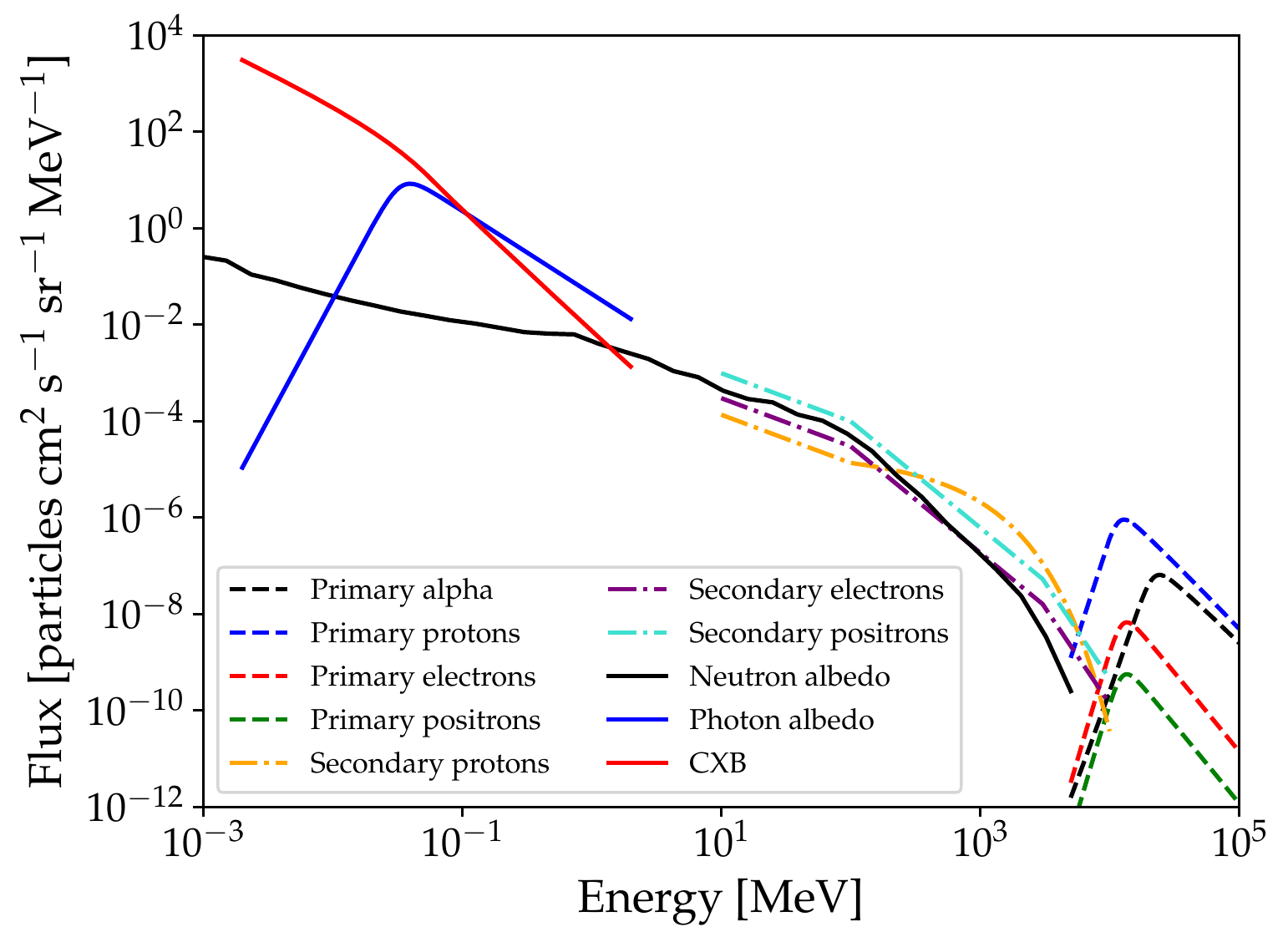}
\caption{Contributions to the THESEUS/XGIS scientific background.}
\label{f:bkg_src}
\end{figure}

The background environment in an equatorial, low-Earth orbit (LEO), has been extensively discussed in other published works\cite{campana13, fioretti12}. In this section we summarize the main populations of particles and photons which contribute to the on-board background, and which are the inputs to the Monte Carlo simulations of the XGIS instrument.
Figure~\ref{f:bkg_src} shows the average differential fluxes of these components.

\subsection{Primary cosmic rays}
For the ``primary'' component of the cosmic ray proton spectrum we assume the unmodulated value given by the BESS and AMS measurements\cite{alcaraz00a, mizuno04}, calculated as a worst case for solar minimum conditions.  
Furthermore, for the other hadronic components of the primary cosmic ray population we consider only the Helium nuclei (i.e. alpha particles),  because the contribution of heavier nuclei is much smaller and in the same range of the uncertainties in the primary flux. 
We consider also the leptonic component of the primary cosmic rays, with a fraction of positrons to electrons that is independent of the energy, i.e. the spectrum of primary positrons has the same slope of the electron one, but a different normalization.
For all these components, we follow the analytical model of Mizuno et al.\cite{mizuno04} for the descriptions of the spectra, with the parameter values corresponding to low geomagnetic latitudes (Figure~\ref{f:bkg_src}). The Earth geomagnetic field cutoff is typically at GeV energies.

\subsection{Secondary particles}
For the low altitude equatorial Earth orbits considered, the impinging proton spectrum (outside the trapped particle belt, i.e. the South Atlantic Anomaly) consists, beside the primary component discussed before, also of a secondary, quasi-trapped component, originating from and impacting to the Earth atmosphere (sometimes referred in the literature as the ``splash'' and ``reentrant'' components). 
The AMS measurements\cite{alcaraz00a} showed that this secondary component is composed of a short-lived and a long-lived particle population, both originating from the regions near the geomagnetic equator.
For low geomagnetic altitudes, the modelling the secondary equatorial proton and lepton spectrum is as a cutoff power-law or a broken power law\cite{campana13, mizuno04} (Figure~\ref{f:bkg_src}).
At variance with respect to the primary particles, in the geomagnetic equatorial region the positrons are predominant with respect to the electrons, with the same overall spectral shape but a ratio e$^{+}$/e$^{-}$ of about 3.3.

There is presently no model for atmospheric albedo neutron fluxes considered mature enough to be used as a standard\footnote{As reported by the official ESA ECSS documents, {\url{http://space-env.esa.int/index.php/ECSS-10-4.html}}}. To account for the flux of neutrons produced by cosmic-ray interactions in the Earth atmosphere, we used the results of the Monte Carlo radiation transport code based QinetiQ Atmospheric Radiation Model (QARM\cite{lei04,campana13}, Figure~\ref{f:bkg_src}).

\subsection{Photon background}

For the cosmic X-ray and $\gamma$-ray diffuse background we assume the Gruber et al.\cite{gruber99} analytic form, derived from HEAO-1 A4 measurements, valid in the range from 3 keV to 100 GeV (Figure~\ref{f:bkg_src}).

The secondary photon background is due to the cosmic-ray (proton and leptonic components) interaction with the Earth atmosphere. As such, it has a strong zenith dependence and a higher flux, for unit of solid angle, than the CXB for energies above 70 keV. 
We assume the parameterized function for the albedo spectrum as given by Swift/BAT\cite{ajello08} measurements (Figure~\ref{f:bkg_src}), that agrees in the range above 50 keV, after some corrections, with previous measurements\cite{campana13}.

\section{THE XGIS MONTE CARLO MASS MODEL}\label{s:mc}
\begin{figure}[htbp]
\centering
\includegraphics[width=6cm]{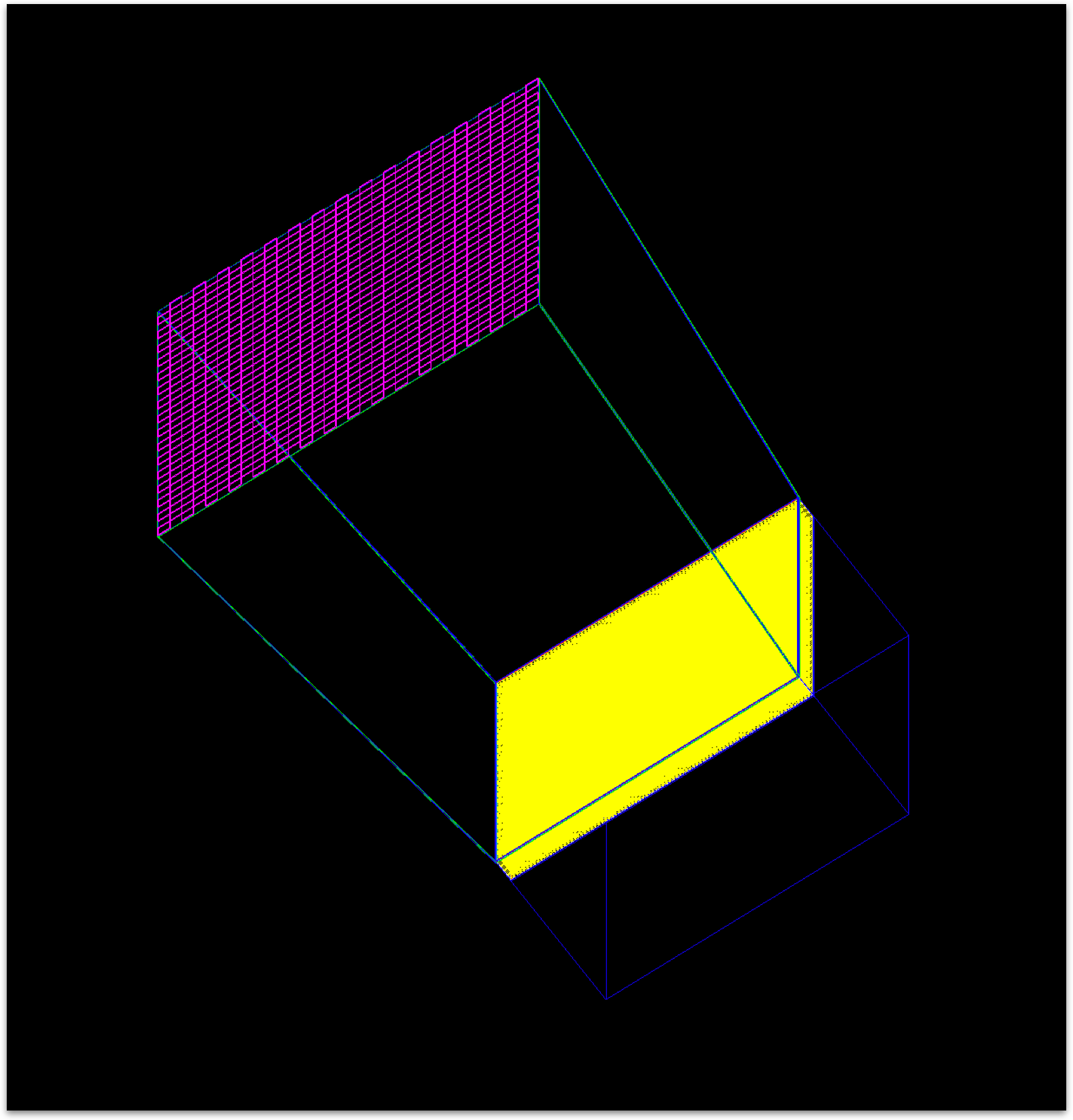}
\caption{Geant-4 mass model of a XGIS unit.}
\label{f:massmodel}
\end{figure}

The Monte Carlo simulator for XGIS is implemented in Geant-4\cite{agostinelli03}. The mass model includes a simplified description of a XGIS unit (Figure~\ref{f:massmodel}) with the following components:
\begin{itemize}
\item Coded mask as a W layer with checkerboard pattern and an open fraction of 50\%.
\item Detection plane with 6400 $4.5\times 4.5\times30$~mm$^3$ CsI(Tl) scintillator bars, coupled to two 450~$\mu$m thick Si layers at both ends (top and bottom SDDs)
\item Side collimator with Al + W layers
\item Front-end and back-end electronics (FEE/BEE) as a simplified Al box
\item Simplified bus structure (Al box with effective density)
\end{itemize}

\begin{figure}[htbp]
\centering
\includegraphics[width=8cm]{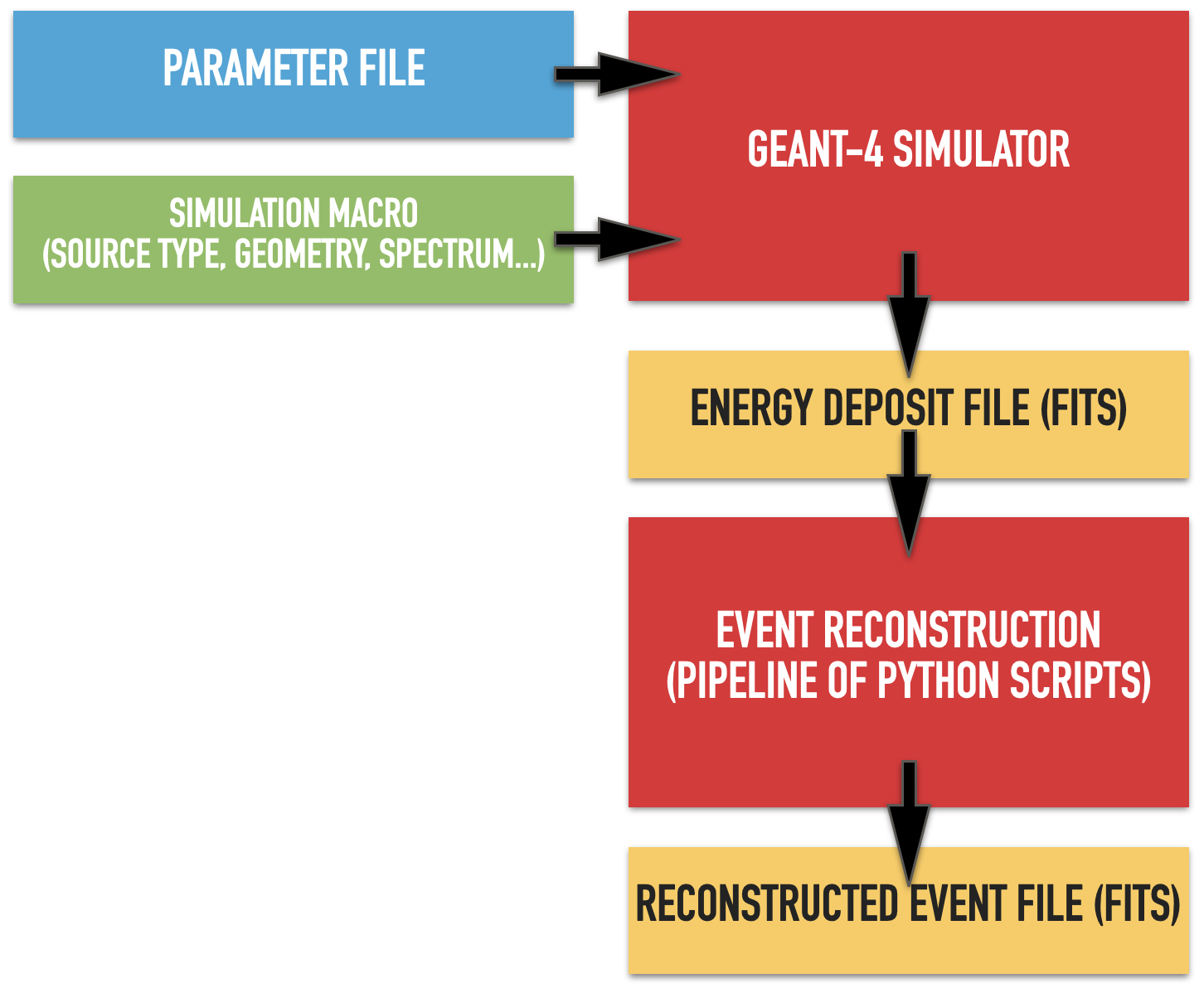}
\caption{The simulation workflow.}
\label{f:workflow}
\end{figure}

All dimensions (thicknesses, relative placements, etc.) and materials are parameterised and can be readout at runtime from an external file. The simulation workflow is shown in Figure~\ref{f:workflow}: the energy deposits inside the Si layers and the CsI bars are registered (in the latter case also as a function of depth), and then an offline event reconstruction that takes into account the scintillation light collection efficiency and the front-end electronic noise is performed using a pipeline of external scripts. The overall simulation output is a FITS file containing the list of each detected events, with all the relevant information (energy, address, primary event information, etc.) for the following derivation of instrumental characteristics.

\section{THE XGIS BACKGROUND AND RESPONSE}\label{s:results}
\subsection{Background}
For the background simulations, the input sources discussed in Section~\ref{s:sources} are simulated isotropically in a spherical surface surrounding the instrument. Effective exposure times are calculated from the input flux and from the detail of the event generation\cite{campana13}. The contribution of each source to the total background is then derived, separately, for the X-mode and for the S-mode.

\begin{figure}[htbp]
\centering
\includegraphics[width=12cm]{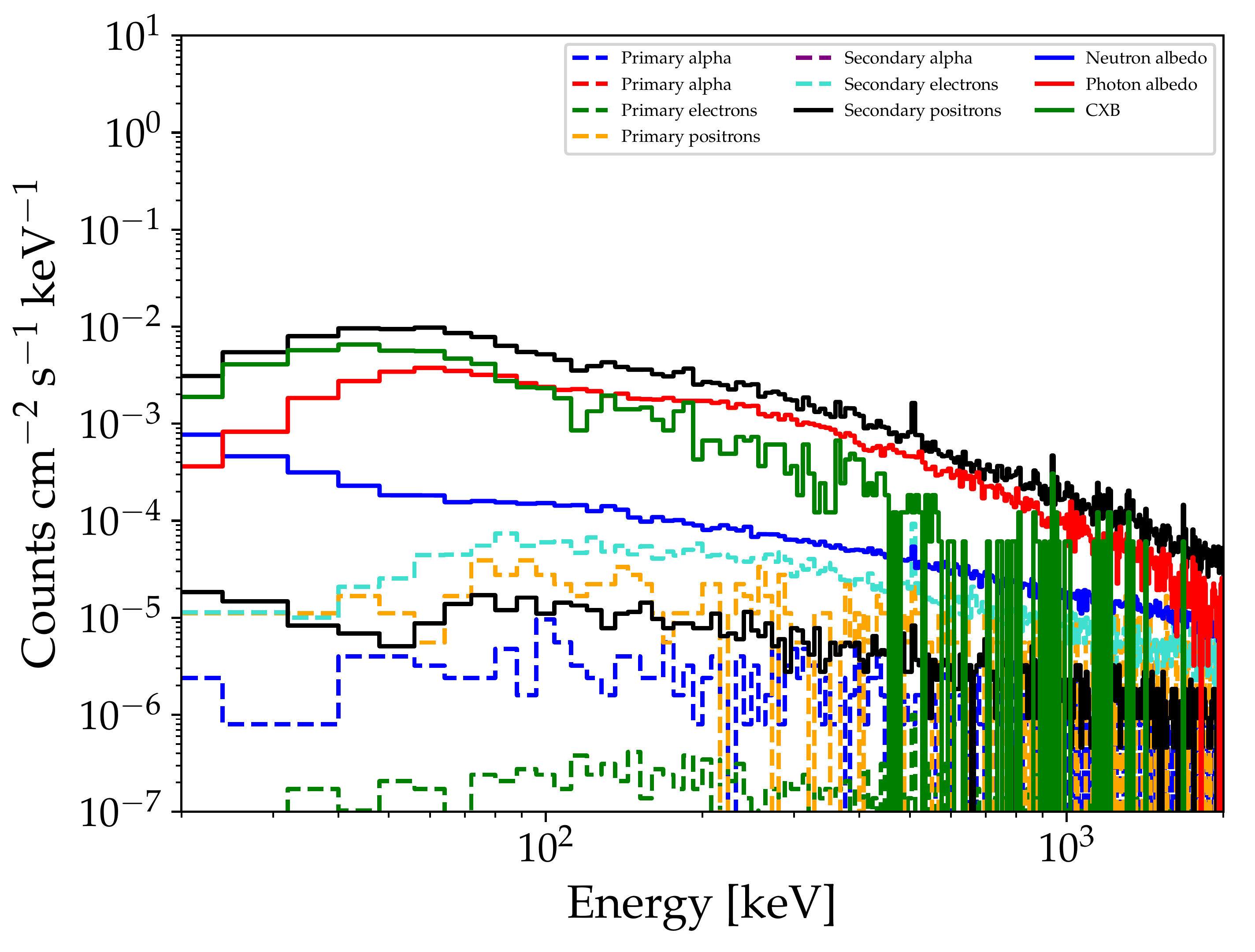}
\caption{Contributions to the THESEUS/XGIS scientific background.}
\label{f:bkg_res}
\end{figure}

For the X-mode, the background is dominated by the cosmic X-ray diffuse radiation collected in the field of view. For the S-mode, instead, photon-induced background (CXB and Earth albedo) dominates at low energies, while at high energies the particles become the main contribution (Figure~\ref{f:bkg_res}).

\begin{figure}[htbp]
\centering
\includegraphics[width=\textwidth]{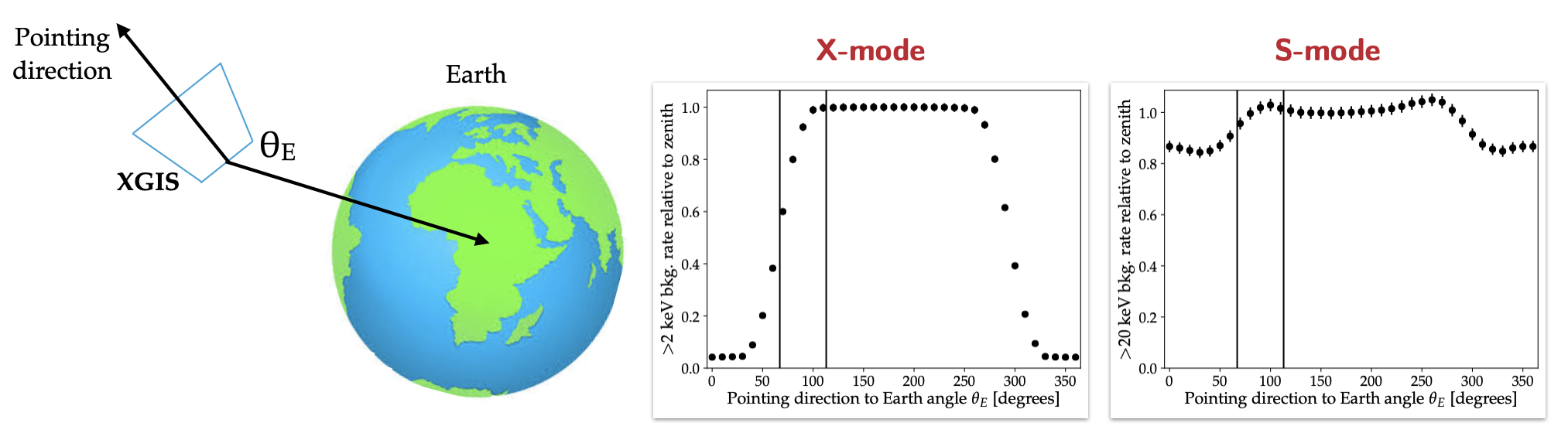}
\caption{\emph{Left panel}: the definition of the angle $\theta_E$. \emph{Central panel}: X-mode total background rate as a function of $\theta_E$, normalised with respect to the case $\theta_E = 180^\circ$, i.e. zenithal pointing.  \emph{Right panel}: the same for S-mode.}
\label{f:theta_E}
\end{figure}

The Earth location with respect to the XGIS unit axis has a significant impact on the resulting background. The Earth blocks about 30\% of the sky at the given orbit: since the CXB and albedo emission have different spectral shapes and fluxes for unit solid angle, the resulting total background is modulated with respect to the pointing direction to the Earth center angle $\theta_E$ (Figure~\ref{f:theta_E}). For the S-mode, the maximum modulation is about 20\%, while for the X-mode, since at low energies the Earth is significantly darker than the sky, the modulation is much higher.

\begin{figure}[htbp]
\centering
\includegraphics[width=8cm]{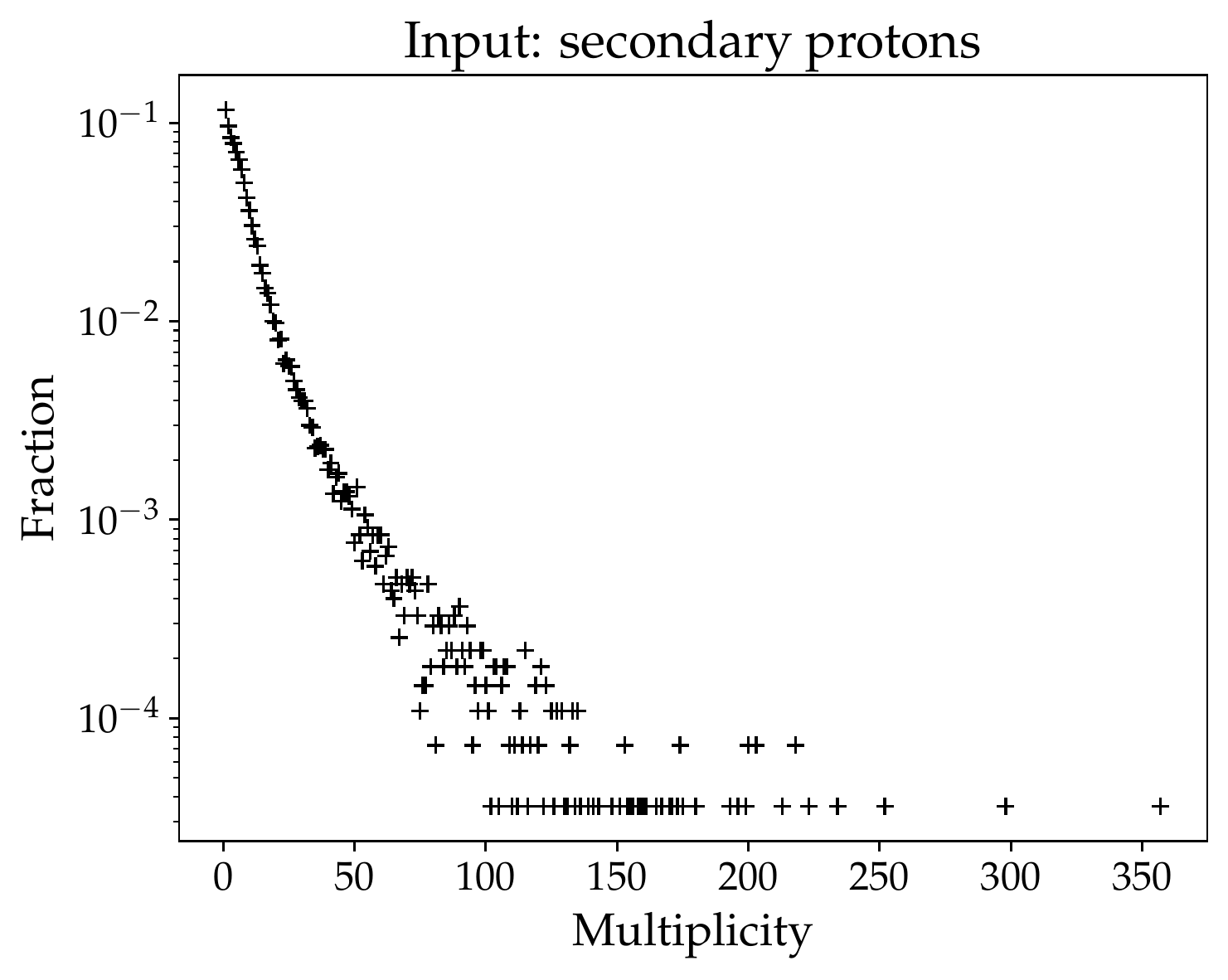}
\caption{Event multiplicity distribution for a input distribution of secondary protons.}
\label{f:mult_secondaryProtons}
\end{figure}

The fine segmentation of the XGIS detection plane (6400 individual pixels) allows not only a certain redundancy, but also to recognize and reject a significant fraction of the particle-induced background. Preliminary simulations show that a \emph{topological} filter that includes both the occurrence of an energy deposit above a few MeV on a single crystal (event saturation) and the number of simultaneously triggering crystals (event multiplicity) has a rejection efficiency up to $\sim$90\%, depending on the input particle type and energy. Figure~\ref{f:mult_secondaryProtons}, for example, shows the event multiplicity resulting from a simulation with secondary protons as input (cf. Section~\ref{s:sources}). A large fraction of the events originating from this population of protons have a multiplicity above 5 (i.e. triggering simultaneously more than five crystals in an ionization streak), while photon-induced events have typical multiplicities below 2--3, with higher multiplicities  possible (e.g. multiple Compton scatterings and/or pair production), but very unlikely and only at the very high end of the sensitivity band.

\subsection{Response}
\begin{figure}[htbp]
\centering
\includegraphics[width=12cm]{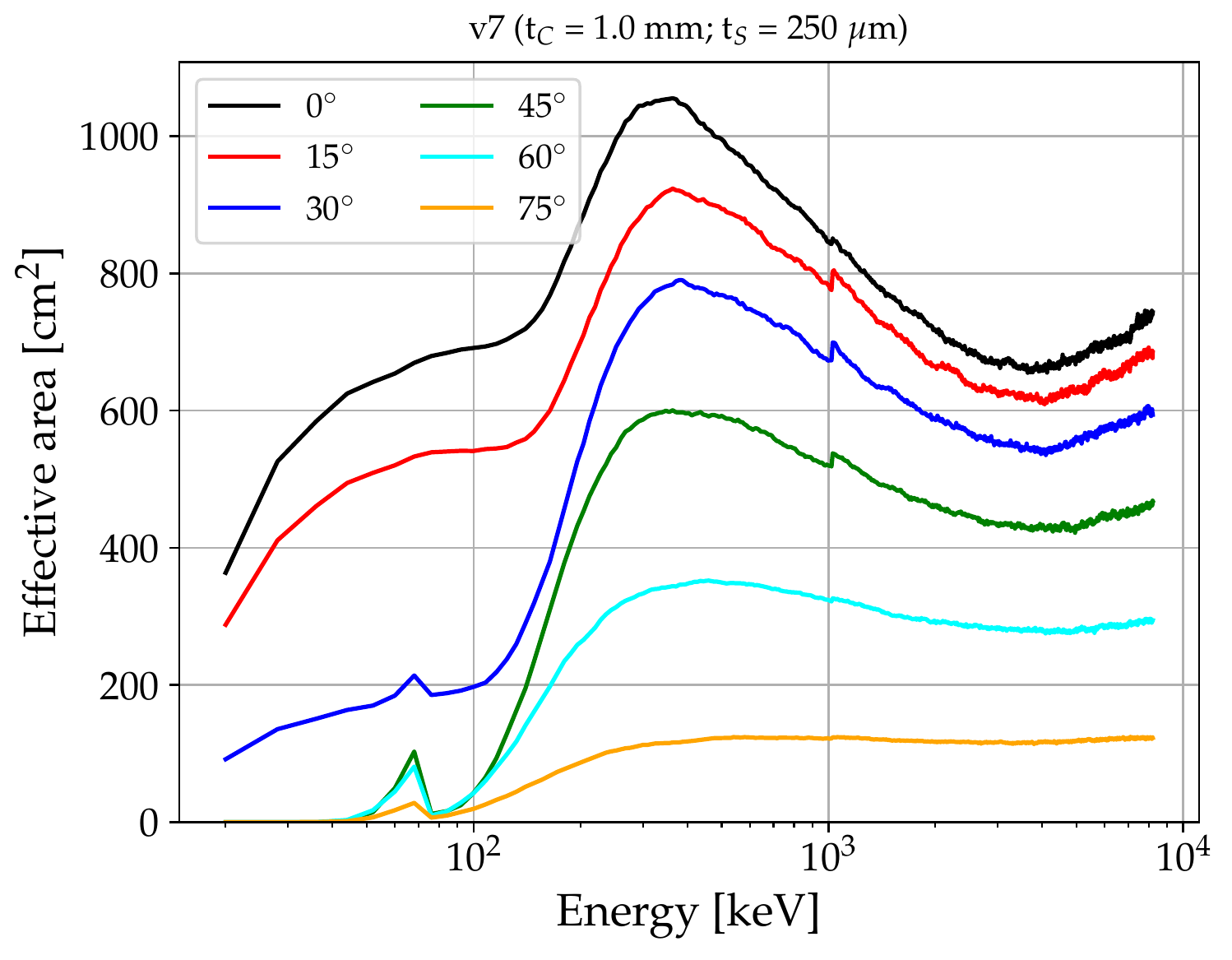}
\caption{S-mode effective area at various off-axis angles.}
\label{f:effarea_offaxis_cmp}
\end{figure}

Another important output of the Monte Carlo simulator is the determination of XGIS response matrices (i.e. the redistribution function and effective area), which is a  fundamental input for scientific simulations\cite{mereghetti20}. The response can be derived by performing a Monte Carlo simulation at several monochromatic energies and at different  angles with respect to the detector axis. Figure~\ref{f:effarea_offaxis_cmp} shows the effective area for the X-mode as a function of the off-axis angles. The present design of the XGIS camera has an efficient collimation in the imaging range (up to $\sim$150~keV), while at higher energies it approaches an isotropic sensitivity (modulated by a pseudo-cosine factor).

\section{CONCLUSIONS AND FUTURE PERSPECTIVES}\label{s:conclusions}
In the framework of the THESEUS ESA M5 Phase A study, a mass model and simulator for the XGIS instrument has been implemented using the Geant-4 toolkit, and a model of the radiative environment in a LEO has been developed, allowing to obtain instrumental background and instrument response files.
The simulator allowed also to study background dependence on instrument attitude (orbital modulation) and the characteristics of the energy deposits (topology).

Future developments of the mass model, depending on the maturity of the XGIS instrument design and of the THESEUS satellite platform development, would include a more detailed mass model for the spacecraft bus and higher level details around the detectors. Moreover, the background simulations should include a study of activation induced background (given the equatorial LEO, preliminary evaluations shows that this should be a minor, albeit non-negligible, contribution to the overall background). The detailed study of event topology, taking into account also the 3D position reconstruction allowed by the dual readout of the scintillator bars, should lead to more efficient background rejection filters thus optimizing the overall sensitivity of the XGIS instrument.

\acknowledgments 
We acknowledge support from the ASI-INAF Agreement n. 2018-29-HH.0, OHB Italia - INAF/OAS Agreement n.2331/2020/01 and ESA Support through the M5/NPMC Programme.
Scientific simulation tools for THESEUS are available at \url{http://www.isdc.unige.ch/theseus/simulation-tools.html}.

\bibliography{report} 
\bibliographystyle{spiebib} 

\end{document}